\documentclass[letter,traditabstract]{aa} 
\usepackage{natbib}
\usepackage{graphicx}
\usepackage{txfonts}
\usepackage{enumitem}

\begin{document} 
\title{Far-infrared properties of submillimeter and optically faint radio galaxies\thanks{Herschel is an ESA space observatory with science instruments provided by European-led Principal Investigator consortia and with important participation from NASA.}}
\author{B. Magnelli\inst{1}
        \and
        D. Lutz\inst{1}      
        \and
        S. Berta\inst{1}
        \and
        B. Altieri\inst{2}
        \and
        P. Andreani\inst{3,4}
        \and
        H. Aussel\inst{5}
        \and
        H. Casta{\~n}eda\inst{6,7}
        \and
        A. Cava\inst{6,7}
        \and
        J. Cepa\inst{6,7}
        \and
        A. Cimatti\inst{8}
        \and
        E. Daddi\inst{5}
        \and
        H. Dannerbauer\inst{5}
        \and
        H. Dominguez\inst{9}
        \and
        D. Elbaz\inst{5}
        \and
        N. F{\"o}rster Schreiber\inst{1}
        \and
        R. Genzel\inst{1}
        \and
        A. Grazian\inst{10}
        \and
        C. Gruppioni\inst{9}
        \and
        G. Magdis\inst{5}
        \and
        R. Maiolino\inst{10}
        \and
        R. Nordon\inst{1}
        \and
        I. P{\'e}rez Fournon\inst{6,7}
        \and
        I. P{\'e}rez Garc{\'\i}a\inst{2}
        \and
        A. Poglitsch\inst{1}
        \and
        P. Popesso\inst{1}
        \and
        F. Pozzi\inst{8}
        \and
        L. Riguccini\inst{5}
        \and
        G. Rodighiero\inst{11}
        \and
        A. Saintonge\inst{1}
        \and
        P. Santini\inst{10}
        \and
        M. Sanchez-Portal\inst{2}
        \and
        L. Shao\inst{1}
        \and
        E. Sturm\inst{1}
        \and
        L. Tacconi\inst{1}
        \and
        I. Valtchanov\inst{5}
        \and
        E.Wieprecht\inst{1}
        \and
        E. Wiezorrek\inst{1}
        }
\offprints{B. Magnelli, \email{Magnelli@mpe.mpg.de}}

\institute{\centering \vskip -10pt \small \it (See online Appendix \ref{sect:affiliations} for author affiliations) }

\date{Received ??; accepted ??}

\abstract{
We use deep observations obtained with the Photodetector Array Camera \& Spectrometer (PACS) onboard the \textit{Herschel} space observatory to study the far-infrared (FIR) properties of submillimeter and optically faint radio galaxies (SMGs and OFRGs).
From literature we compiled a sample of 35 securely identified SMGs and nine OFRGs located in the GOODS-N and the A2218 fields.
This sample is cross-matched with our PACS 100~$\mu$m and 160~$\mu$m multi-wavelength catalogs based on sources-extraction using prior detections at 24 $\mu$m.
About half of the galaxies in our sample are detected in at least the PACS 160 $\mu$m bandpass.
The dust temperatures and the infrared luminosities of our galaxies are derived by fitting their PACS and SCUBA 850~$\mu$m (only the upper limits for the OFRGs) flux densities with a single modified ($\beta=1.5$) black body function.
The median dust temperature of our SMG sample is $T_{dust}=36\pm8$~K while for our OFRG sample it is $T_{dust}=47\pm3$~K.
For both samples, median dust temperatures derived from \textit{Herschel} data agree well with previous estimates.
In particular, Chapman et al. (2005) found a dust temperature of $T_{dust}=36\pm7$~K  for a large sample of SMGs assuming the validity of the FIR/radio correlation (i.e., $q=log_{10}(L_{FIR}[{\rm W}]/L_{1.4\,{\rm GHz}}[{\rm W\,Hz^{-1}}]/3.75\times10^{12})$).
The agreement between our studies confirms that the local FIR/radio correlation effectively holds at high redshift even though we find $\langle q\rangle=2.17\pm0.19$, a slightly lower value than that observed in local systems.
The median infrared luminosities of SMGs and OFRGs are $4.6\times10^{12}\,L_{\odot}$ and $2.6\times10^{12}\,L_{\odot}$, respectively.
We note that for both samples the infrared luminosity estimates from the radio part of the spectral energy distribution (SED) are accurate, while estimates from the mid-IR are considerably ($\thicksim\times3$) more uncertain.
Our observations confirm the remarkably high luminosities of SMGs and thus imply median star-formation rates of 960~M$_{\odot}\,$yr$^{-1}$ for SMGs with $S(850\,\mu{\rm m})>5\,$mJy and 460~M$_{\odot}\,$yr$^{-1}$ for SMGs with $S(850\,\mu{\rm m})>2\,$mJy, assuming a Chabrier IMF and no dominant AGN contribution to the far-infrared luminosity.
}
\keywords{Galaxies: evolution - Infrared: galaxies - Galaxies: starburst - Submillimeter: galaxies}
\authorrunning{Magnelli et al. }
\titlerunning{Far-infrared properties of SMGs and OFRGs}
\maketitle

\section{Introduction}
\indent{
\textit{Herschel} observations probe the rest-frame far-infrared emission of high-redshift galaxies.
Thus, they provide for the first time robust estimates of the infrared luminosities of these high-redshift galaxies and test previous measurements that were based on extrapolation from shorter or longer wavelengths.
We here  focus on two populations of high-redshift star-forming galaxies selected at submillimeter (submm) and radio wavelengths.
\\}
\indent{
Since their discovery in the late 1990s, submillimeter galaxies (SMGs) have become the selection of choice for the most luminous tail of the high-redshift star-forming galaxy population.
It has been found that SMGs have a typical redshift of $z\thicksim2$ \citep{chapman_2005,pope_2006}, are massive systems \citep[$M_{\star}\thicksim10^{10}-10^{11}\,M_{\odot}$,][]{swinbank_2004,tacconi_2006} and are compact \citep[e.g.,][]{tacconi_2008}.
Interferometric observations of their CO molecular gas suggest that the most luminous SMGs ($S_{850\,\mu {\rm m}}> 5$ mJy) are merging sytems \citep{tacconi_2006,tacconi_2008} with high star-formation efficiencies compared to typical galaxies of a similar mass \citep{daddi_2008}.
Therefore, these SMGs are thought to exhibit very intense (SFR$\thicksim$1000 M$_{\odot}$ yr$^{-1}$) short-lived star-formation bursts triggered by mergers and to be the high-redshift progenitors of local massive early-type galaxies \citep{daddi_2007,daddi_2007a,tacconi_2008,cimatti_2008}. 
\\}
\indent{
Although SMGs provide a powerful tool to constrain the formation and the evolution of high-redshift dusty star-forming galaxies, their selection is subject to strong biases.
In particular, because submm observations probe the blackbody emission of dust in the Rayleigh-Jeans regime, they are strongly anti-correlated with the dust temperature ($S_{850}\propto T^{-3.5}_{dust}$).
For a given infrared luminosity, galaxies with hot dust might fall below the detection limit of current submm instruments.
The first observational evidence of a missing population of high-redshift dusty star-forming galaxies with hot dust has been  given by \citet{chapman_2004} using a selection of radio-detected but submm-faint galaxies with UV spectra consistent with high-redshift starbursts.
These optically faint-radio galaxies (OFRGs) have a comoving volume density (i.e., $\thicksim10^{-5}\,$Mpc$^{-3}$ at $1<z<3$, Chapman et al. \citeyear{chapman_2004}), stellar masses and sizes comparable to SMGs, and some have a dust temperature of $\thicksim52~$K \citep{casey_2009a,casey_2009b}.
\\}
\indent{
While SMGs and OFRGs are an important component of the high-redshift massive galaxy population, many of their fundamental properties still rely on indirect measurements.
In particular, direct determinations of SMG dust temperatures were limited \citep{kovacs_2006} because they were not done using rest-frame far-infrared observations on both sides of the peak of the SEDs.
More importantly, their infrared luminosites are still debated.
Indeed, because theoretical simulations of galaxy evolution have had great difficulties to account for the current inferred luminosities/star-formation rates and number counts \citep{baugh_2005,dave_2010}, they still question whether these luminosities have been overestimated or whether the IMF is significantly more top-heavy than in the local Universe. 
\\}
\indent{
Using deep observations by the Photodetector Array Camera \& Spectrometer (PACS; Poglitsch et al. \citeyear{poglitsch_2010}) onboard the \textit{Herschel} space observatory (Pilbratt et al. \citeyear{pilbratt_2010}) we will have for the first time robust estimates of the dust temperatures and the infrared luminosities of SMGs and OFRGs.
Throughout the paper we use a cosmology with $H_{0}=70 \rm{km\,s^{-1}\,Mpc^{-1}}$, $\Omega_{\Lambda}=0.7$ and $\Omega_{\rm M}=0.3$.
\\}
\begin{figure*}
\centering
         \includegraphics[width=12.cm]{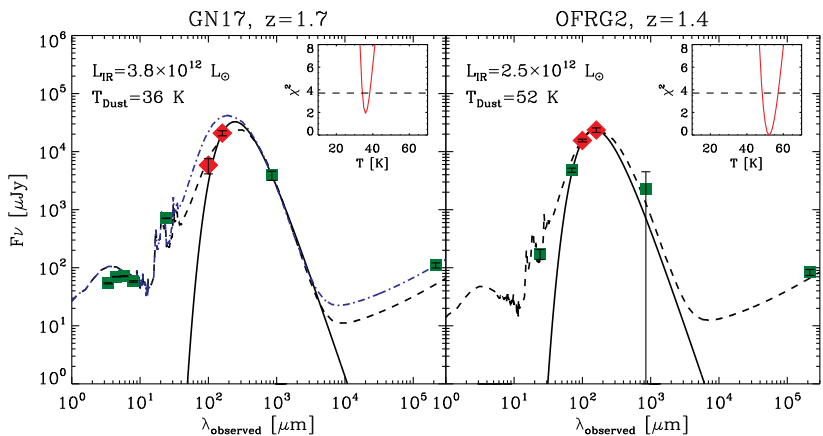}
	\caption{\label{fig: fit BB}\small{
	Spectral energy distribution (SED) of one SMG (\textit{left}) and one OFRG (\textit{right}).
	Red diamonds present our PACS measurements while green squares present the multi-wavelength ancillary data taken from the literature \citep{pope_2006,casey_2009a,casey_2009b}.
	The modified blackbody emission ($\beta=1.5$) best-fitting the data are shown by solid lines.
	Dashed lines present the  Dale \& Helou SED template best-fitting the mid- to far-infrared observations.
	In the left panel, the blue dotted-dashed line shows the Dale \& Helou SED template best-fitting the submm and radio photometries.	
	The inset in each panel shows $\chi^{2}$ vs $T_{dust}$.
	}}
\end{figure*}
\section{Observations}
\indent{
In this study we used deep PACS 100 $\mu$m and 160 $\mu$m observations of the Great Observatories Origins Deep Survey-North (GOODS-N; $12^{h}36^{m},\,+62^{\circ}14\arcmin$) and the Abell 2218 ($16^{h}35^{m},\,+66^{\circ}12\arcmin$) fields.
These observations were taken as part of the PACS Evolutionary Probe (PEP\footnote{http://www.mpe.mpg.de/ir/Research/PEP}) guaranteed time key program.
The GOODS-N field covers a region of $10\arcmin\times15\arcmin$ (30 hours), while the deep part of the A2218 field covers a region of $4\arcmin\times4\arcmin$ (13 hours).
\\}
\indent{
At the resolution of \textit{Herschel}, all sources in our fields are point sources (i.e. FWHM$\,\thicksim8\arcsec$ [$12\arcsec$] at 100 $\mu$m [160~$\mu$m]).
Flux densities are hence estimated using a point spread function-fitting technique based on prior source positions detected at 24~$\mu$m.
The use of priors provides a straightforward association between the IRAC, MIPS and PACS sources.
Using Monte Carlo simulations we estimate the quality of our PACS 100~$\mu$m and 160~$\mu$m catalogs, i.e. photometric error, completeness and contamination as a function of the flux density.
In the GOODS-N field our observations reach a $3\sigma$ limit of $\thicksim3$~mJy and $\thicksim5.7$~mJy at 100~$\mu$m and 160~$\mu$m respectively, while in the A2218 field they reach a $3\sigma$ limit of $\thicksim2.5$~mJy and $\thicksim5$~mJy at 100~$\mu$m and 160~$\mu$m respectively.
\\}
\indent{
A complete description of PEP data reduction and sources extraction is given in Appendix A of Berta et al. (\citeyear{berta_2010}).
\\}
\section{Galaxy sample}
\indent{
To obtain a robust measurement of the dust temperature and infrared luminosity of a given galaxy one needs to have an accurate estimate of its redshift.
Consequently, we decided to restrict our study to a sample of SMGs and OFRGs with accurate redshift estimates derived from secured radio/mid-infrared identifications (PACS identifications of SMGs are presented in Dannerbauer et al. in prep).
In the GOODS-N field, our SMG sample is based on multi-wavelength identifications of SCUBA and AzTEC sources made by \citet{pope_2006}\footnote{For GN05, GN07, GN10, GN20 and GN20.2, we used the spectroscopic redshifts revised in \citet{pope_2008} and \citet{daddi_2009b,daddi_2009a}.} and \citet{chapin_2009}, respectively.
SMGs with tentative redshifts determined from their IRAC or mid/far-infrared/radio colors were excluded from our sample.
Sources with multiple optical counterparts (GN04, GN07, GN19 and GN39) were treated as a single system (i.e. we will use the sum of the radio and mid-infrared flux from the two components when determining their far-infrared properties) because they are all thought to be interacting galaxies \citep{pope_2006}. 
All these different criteria yield an SMG sample containing 29 sources in the GOODS-N field.
In the A2218 field, our SMG sample is assembled from the literature \citep{kneib_2004,knudsen_2006,knudsen_2008} and contains six lensed sources.
Because these galaxies are magnified, their mid-to-far infrared fluxes were de-magnified prior to further analysis using magnification factors from the above references.
Among these six lensed sources, three correspond to the same lensed galaxy \citep[SMMJ16359+6612; ][]{kneib_2004}.
Finally, our OFRG sample is taken from \citet{casey_2009a,casey_2009b} and contains nine sources, all situated in the GOODS-N field.
We note that all but three sources of our entire sample (i.e. SMGs and OFRGs) have spectroscopic redshifts.
\\}
\indent{
The SMG and OFRG samples were cross-matched with our PACS multi-wavelength catalogs using a matching radius of $3\arcsec$.
We detected 19 out of 35 SMGs in at least the PACS 160 $\mu$m bandpass (17 out of 33 if not multi-counting the 3-component lensed source detected in A2218).
The PACS sample is slightly biased towards lower redshift sources because of the positive \textit{K-}correction : while the median redshift of our parent SMG sample is $z=2$, the median redshift of our PACS detected SMG sample is $z=1.7$.
Five out of nine OFRGs have PACS 100~$\mu$m and 160~$\mu$m detections.
This sample is also slightly biased toward lower redshift ($z=1.5$) because the OFRG situated at the highest redshift is undetected in our PACS images.
\\}
\indent{
We note that our SMG sample contains sources with $2\,{\rm mJy}<S_{850\,\mu{\rm m}}<5\,{\rm mJy}$, while the most luminous tail of the SMG population, mostly associated with major mergers, is defined using $S_{850\,\mu{\rm m}}>5\,{\rm mJy}$.
Below, we will draw our conclusions distinguishing these two populations of SMGs.
\\}
\begin{figure}
\centering
	\includegraphics[width=7.5cm]{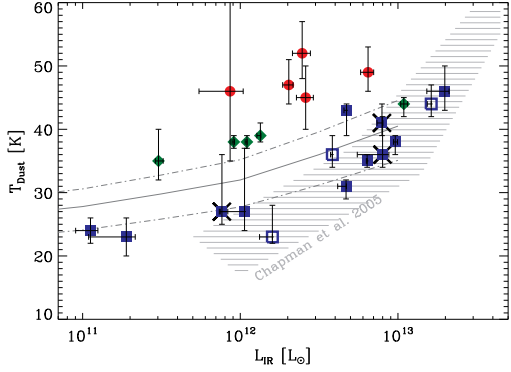}
	\caption{\label{fig: t vs l}\small{
	Dust temperature-luminosity relation.
	The filled blue squares and opened blue squares denote SMGs located in the GOODS-N fields with spectroscopic and photometric redshift, respectively.
	The crosses denote sources which contain an AGN, as indicated by the presence of hard X-rays.
	The filled green diamonds present SMGs located in the A2218 field (the three data points with $L_{IR}\thicksim1\times10^{12}\,L_{\odot}$ correspond to the same lensed galaxy).
	The filled red circles are from our OFRGs sample.
	The striped area presents results for SMGs extrapolated by \citet{chapman_2005} from radio and submm data.
	The \citet{chapman_2003} derivation of the median and interquartile range of the $T_{dust}$-$L_{IR}$ relation observed at $z\thicksim0$ is shown by solid and dashed-dotted lines, linearly extrapolated to $10^{13}\,L_{\odot}$.
	}}
\end{figure}
\section{Data analysis}
\indent{
In order to infer the dust temperature of our galaxies we fitted their PACS and SCUBA photometry (only the upper limit for the OFRGs) with a modified blackbody function, with a dust emissivity $\beta=1.5$ (see Fig \ref{fig: fit BB}).
Their total infrared luminosities ($L_{IR}$[8-1000\,$\mu$m]) were inferred from these best fits using the far-infrared luminosity definition ($L_{FIR}[40-120\,\mu$m]) given by Helou et al. (\citeyear{helou_1988}) and a color-correction term \citep[][$L_{IR}=1.91\times L_{FIR}$]{dale_2001}.
\\}
\begin{figure}
\centering
	\includegraphics[width=7.5cm]{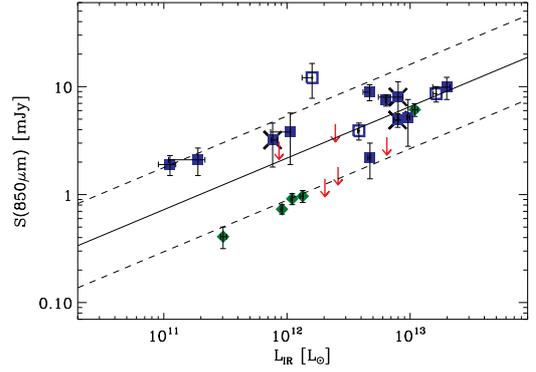}
	\caption{\label{fig: s850 vs lir}\small{Submm flux densities as function of the infrared luminosity.
	The symbols are same as in Fig \ref{fig: t vs l} except for the OFRGs, for which we have only upper limits.
	The solid and dashed lines show the linear fit to the $S_{850\,\mu{\rm m}}$-$L_{IR}$ relation and the 1$\sigma$ envelop ($L_{IR}[L_{\odot}]=10^{11.3\pm0.8}\times S_{850\,\mu{\rm m}}^{2.1\pm0.1}\,$[mJy]).
	If we remove the lensed-SMGs of A2218 from the fit, we find a weaker correlation ($L_{IR}[L_{\odot}]=10^{10.0\pm0.4}\times S_{850\,\mu{\rm m}}^{3.5\pm0.1}\,$[mJy]).
	}}
\end{figure}
\indent{
We adopted a single dust temperature characterization because studies of IRAS galaxies have demonstrated that this provides an accurate diagnostic of the typical heating condition in their interstellar medium \citep{desert_1990}.
While for most of our galaxies this single dust temperature characterization provides a good description of their far-infrared SED, for 6 SMGs, this single dust temperature model yields high $\chi^2$ values (i.e. $\chi^{2}>2.71+N_{dof}$).
All these galaxies appear either to be the more distant ones or to exhibit far-infrared colors typical of very cold systems.
In both cases, their PACS 100 $\mu$m flux densities might by contaminated by a hotter dust component.
Therefore we excluded their PACS 100~$\mu$m photometry from the fit and recomputed their dust temperatures.
We note that excluding the PACS 100~$\mu$m photometry from the fit of all our galaxies changes their median dust temperature by only $\Delta T_{dust}\thicksim-2$ K.
\\}
\indent{
We tested that our results are relatively insensitive to $\beta$ and to our single dust component characterization.
Indeed we note that using $\beta=2$, we found only small differences in the values of $T_{dust}$ ($\Delta T_{dust}\thicksim+3$ K).
Moreover,  we also note that to fit the PACS and SCUBA photometry with the multiple dust components model of Dale \& Helou (\citeyear{dale_2002}) yields rest-frame infrared colors ($S_{60\,\mu{\rm m}}/S_{100\,\mu{\rm m}}$), or equivalently $T_{dust}$, which excellently agree with those inferred from our blackbody analysis.
\\}
\section{Discussion}
\indent{
Figure \ref{fig: t vs l} depicts the locations of our SMGs and OFRGs on the $T_{dust}$-$L_{IR}$ plane and Fig \ref{fig: s850 vs lir} shows their locations on the $S_{850\,\mu{\rm m}}$-$L_{IR}$ plane.
As already mentioned, OFRGs are biased towards hot dust temperatures; their median $T_{dust}$ is of $47\pm3$~K and their median $L_{IR}$ is $2.6\times10^{12}\,L_{\odot}$.
 In contrast, SMGs have lower dust temperatures with median $T_{dust}=36\pm8$~K and $L_{IR}=4.6\times10^{12}\,L_{\odot}$.
We note that lensed-SMGs from A2218 and with $L_{IR}<2\times10^{12}\,L_{\odot}$ exhibit intermediate dust properties and are less biased towards cold dust temperatures than the entire SMG sample.
This is because these galaxies would have escaped both the SMG and OFRG selection method without magnification.
We also note that \textit{bright} SMGs (i.e. $S_{850\,\mu{\rm m}}>5\,{\rm mJy}$) have higher median infrared luminosities ($L_{IR}=9.6\times10^{12}\,L_{\odot}$) and higher median dust temperatures ($T_{dust}=38$ K) than the entire SMG sample because there is a correlation between $S_{850\,\mu{\rm m}}$ and $L_{IR}$ (Fig \ref{fig: s850 vs lir}).
These estimates are the first direct observational measurements of the dust temperatures and the infrared luminosities of SMGs and OFRGs.
\\}
\indent{
Our observations reveal that high redshift dusty star-forming galaxies exhibit a wide range of dust temperatures.
In particular at low infrared luminosities ($L_{IR}<4\times10^{12}L_{\odot}$) the dust temperature dispersion observed in our sample might suggest a higher $T_{dust}$-$L_{IR}$ scatter than that observed by \citet{chapman_2003} at $z\thicksim0$.
Nevertheless, this conclusion is most likely driven by selection effects because a significant fraction of the galaxies with intermediate dust properties were probably missed by our current sample.
Indeed, we note that studying a $L_{IR}$-selected sample of galaxies observed with \textit{Herschel}, Hwang et al. (in prep) find modest changes in the $T_{dust}$-$L_{IR}$ relation as function of the redshift :  at $z>0.5$, galaxies with $L_{IR}>5\times10^{10}L_{\odot}$ are slightly colder ($\thicksim3\,$K) than local ones and the scatter of the $T_{dust}$-$L_{IR}$ relation slightly increase at high redshift.
\\}
\indent{
Though previous estimates of the dust temperatures of SMGs and OFRGs relied on indirect observations, they agree relatively well with our measurements.
In particular, \citet{chapman_2005} found a dust temperature of $T_{dust}=36\pm7$~K for a large sample of SMGs assuming the validity of the FIR/radio correlation.
In order to establish this agreement on a common sample, we applied the same method as \citet{chapman_2005} to our SMG sample, i.e. we fitted the radio and 850~$\mu$m photometries with dust SED templates from \citet{dale_2002} and then translated them into $T_{dust}$ using their R(60,100) to $T_{dust}$ map.
With this method we found higher $T_{dust}$ (by $\thicksim4\,$K) and $L_{IR}$ ($\thicksim\times1.5$ times) than what we obtained using our blackbody analysis.
These discrepancies arise because the Dale \& Helou SED templates assume a FIR/radio correlation with $\langle q\rangle=2.34$ \citep{yun_2001}, while in our samples we find $\langle q\rangle=2.17\pm0.19$ (see also Ivison et al. \citeyear{ivison_2010b}, this issue).
Although this value of $\langle q\rangle$ is still in line with results from local systems (which have a dispersion of 0.19~dex), it is also consistent with an evolution of $\langle q\rangle$ proportional to $(1+z)^{-0.15\pm0.03}$ as found by \citet{ivison_2010}.
\\}
\indent{
Our results reveal that one can obtain a very reliable estimate of the infrared luminosity of a given galaxy from its radio flux density ($\sigma[log_{10}(L_{IR}^{Radio}/L_{IR}^{Blackboby})]\thicksim0.18$ dex) using $\langle q\rangle=2.17$.
In contrast, the use of the 24 $\mu$m emission and of the Chary \& Elbaz SED library yields an inaccurate estimate of the infrared luminosity characterized by a large scatter ($\sigma[log_{10}(L_{IR}^{24\,\mu m}/L_{IR}^{blackbody})]\thicksim0.48$~dex) and a systematic overestimation ($\thicksim\times2$ times) of the most luminous galaxies ($L_{IR}>4\times10^{12}L_{\odot}$).
We thus find that for the very luminous infrared galaxies studied here, luminosity extrapolations based on the radio emission are considerably more reliable than those based on the mid-infrared emission (see also Elbaz et al. \citeyear{elbaz_2010} and Nordon et al. \citeyear{nordon_2010}).
\\}
\indent{
Using $\langle q\rangle=2.17$, we can predict the infrared luminosities of our PACS undetected SMGs.
Then, using the Dale \& Helou SED templates normalized to these infrared luminosities, we can fit their 850~$\mu$m photometries.
We find that for 13 out of 16 undetected SMGs, PACS flux densities inferred using these fits are below the detection threshold of our observations.
Of the three sources with PACS fluxes predictions above our detection threshold, one is known to be contaminated by an AGN and the other two are suspected to have wrong redshift estimates \citep{daddi_2009a}.
The PACS nondetections are thus fully consistent with the properties inferred from the detections.
This analysis cannot be performed on our PACS undetected OFRGs because they are also undetected at the SCUBA wavelength.
\\ \\}
\indent{
Our observations unambiguously confirm the remarkably large infrared luminosities of \textit{bright} SMGs (i.e. $S_{850\,\mu{\rm m}}>5\,{\rm mJy}$) which correspond to SFRs of $960\ {\rm M}_{\odot}~ {\rm yr}^{-1}$ (SFR~$[{\rm M}_{\odot}~ {\rm yr}^{-1}] = 1\times 10^{-10} L_{\rm IR}~[{\rm L}_{\odot}]$, assuming a Chabrier IMF and no dominant AGN contribution to the far-infrared luminosity).
Such high SFRs are difficult to reconcile with secular evolution \citep[e.g.][]{dave_2010} and could correspond to a merger-driven stage in the evolution of these galaxies.
This hypothesis is supported by CO observations of \textit{bright} SMGs which have revealed large CO line-widths and disturbed gas morphologies \citep{tacconi_2008}.
Our observations also confirm that OFRGs exhibit higher dust temperatures than \textit{faint} SMGs ($S_{850\,\mu{\rm m}}<5\,{\rm mJy}$) observed at the same redshift and with equivalent infrared luminosities.
While the relatively low median SFRs of OFRGs and \textit{faint} SMGs ($260\ {\rm M}_{\odot}~ {\rm yr}^{-1}$ and $106\ {\rm M}_{\odot}~ {\rm yr}^{-1}$ respectively) could be explained by secular evolution, we need to understand why they exhibit different dust temperatures and to study a possible link with 	\textit{bright} SMGs.
A clear evolutionary picture will require detailed studies of the dust and molecular gas distribution in a sample of high-redshift star-forming galaxies unbiased towards any particular dust temperature.
This SFR-selected sample can now be built with our ongoing deep \textit{Herschel} observations.
\\}
\acknowledgement{
PACS has been developed by a consortium of institutes led by MPE (Germany) and including UVIE (Austria); KU Leuven, CSL, IMEC (Belgium); CEA, LAM (France); MPIA (Germany); INAF-IFSI/OAA/OAP/OAT, LENS, SISSA (Italy); IAC (Spain). This development has been supported by the funding agencies BMVIT (Austria), ESA-PRODEX (Belgium), CEA/CNES (France), DLR (Germany), ASI/INAF (Italy), and CICYT/MCYT (Spain).
}

 \Online
\begin{appendix}
\section{Authors' affiliations}\label{sect:affiliations}

\begin{enumerate}[label=$^{\arabic{*}}$]
\item Max-Planck-Institut f\"{u}r Extraterrestrische Physik (MPE), Postfach 1312, 85741 Garching, Germany.
\item Laboratoire AIM, CEA/DSM-CNRS-Universit{\'e} Paris Diderot, IRFU/Service
d'Astrophysique,
B\^at.709, CEA-Saclay, 91191 Gif-sur-Yvette Cedex, France.
\item Herschel Science Centre; European Space Astronomy Centre
\item ESO, Karl-Schwarzschild-Str. 2, D-85748 Garching, Germany.
\item INAF - Osservatorio Astronomico di Trieste, via Tiepolo 11, 34143 Trieste, Italy.
\item Instituto de Astrof{\'i}sica de Canarias, 38205 La Laguna, Spain.
\item Departamento de Astrof\'isica, Universidad de La Laguna, Spain.
\item Dipartimento di Astronomia, Universit{\`a} di Bologna, Via Ranzani 1,
40127 Bologna, Italy.
\item INAF-Osservatorio Astronomico di Bologna, via Ranzani 1, I-40127 Bologna,
Italy.
\item INAF - Osservatorio Astronomico di Roma, via di Frascati 33, 00040 Monte
Porzio Catone, Italy.
\item Dipartimento di Astronomia, Universit{\`a} di Padova, Vicolo
dell'Osservatorio 3,
35122 Padova, Italy.
\end{enumerate}

\end{appendix}

\end{document}